\documentstyle[prd,preprint,aps]{revtex} 

\newcommand{\be}{\begin{equation}}
\newcommand{\ee}{\end{equation}}
\newcommand{\bea}{\begin{eqnarray}}

\newcommand{\eea}{\end{eqnarray}}

\begin{document}

\reversemarginpar
\tighten

\title{A Commentary on Ruppeiner Metrics for Black Holes}
\author{A.J.M.  Medved} 
\address{Physics Department,\\
 University of Seoul, \\
Seoul 130-743, \\
Korea \\
E-mail(1): allan@physics.uos.ac.kr \\
E-mail(2): joey\_medved@yahoo.com \\
}
\maketitle
\begin{abstract}

\par
There has been some recent controversy regarding the  Ruppeiner metrics
 that are induced by
 Reissner--Nordstrom (and  Reissner--Nordstrom-like) black holes. 
Most infamously, why does
this family of metrics turn out to be flat, 
how is this outcome to  be physically understood, and can/should the 
formalism  be suitably  modified to induce curvature? In the current
 paper, we 
provide a novel interpretation of this debate. For the sake of maximal
analytic clarity and tractability,  some supporting calculations are
 carried
out for the relatively  simple model of a rotating BTZ black hole.
\end{abstract}
\newpage
\section{Background}
\par
From a historical perspective (with attention to the  early
  seventies),
the paradigm of black hole thermodynamics can  be viewed as the
synthesis of three main theoretical inputs: {\it (i)} a stunning
 analogy
between black hole mechanics and the ``standard''
laws of thermodynamics \cite{analogy},  {\it (ii)} Bekenstein's
proposal of a ``generalized'' second law of thermodynamics
 \cite{beken} and {\it (iii)} Hawking's  quantum-field-based 
calculation of the  black hole radiation spectrum \cite{hawk}.
(Note that the second point implies that a black hole should
be assigned an entropy, while  the third directly  ascribes it
with a temperature.) Given its rigorous nature, the Hawking calculation
 was
 particularly significant in legitimizing the overall picture. 
Moreover, Hawking's result allowed an unambiguous calibration
for the  thermodynamic properties of interest.

 Now being more
explicit, let us focus on a 
(``non-exotic''~\footnote{Meaning that we are dismissing the 
possibility of any (so-called) ``quantum 
hair'' and other esoteric considerations from the present discussion.})
 black hole of mass $M$, charge 
$Q$ and angular momentum $J$
 in a four-dimensional asymptotically 
flat spacetime. (Also known as a Kerr--Newman black hole.)
 One would find it natural to assign this object with 
 an entropy  ($S$), a temperature ($T$),
an electrostatic potential ($\Phi$) and a  
rotational velocity ($\Omega$) 
 in
 precisely the following manner:
\be
S\;=\;{A\over 4}\;=\; 8\pi M\left[M -{Q^2 \over 2M}
+\sqrt{M^2-Q^2-{J^2\over M^2}}\right] \; ,
\label{1}
\ee
\be
T\;=\;{2\over A}\sqrt{M^2-Q^2-{J^2\over M^2}} \; ,
\label{2}
\ee
\be
\Phi\;=\; Q \sqrt{4\pi\over A}
\label{3}
\ee
and
\be
\Omega \;=\; 4\pi{J\over MA} \;.
\label{4}
\ee
Here, we have set all fundamental constants equal
to unity (these can easily be restored by way
of dimensional analysis) and introduced the notion
of a  black hole having an event horizon with a surface area of $A$.
Actually, black hole thermodynamics is essentially 
based on the properties of this outer-most horizon surface,
and we note, in passing, that $T=\kappa/2\pi$ where
$\kappa$ is the surface gravity 
(which measures the strength of the gravitational
field in the proximity of this special surface).   

\par
As alluded to by point {\it (i)} above, 
the concept of black holes as thermodynamic
 systems
follows --- in part --- from the following quantitative statement:
\be
dM\;=\; TdS \;+\; \Phi dQ \; + \; \Omega dJ \;,
\label{5}
\ee
which is clearly analogous to the first law of thermodynamics. 
But, strictly speaking, this analogy is only  provisional:  it depends 
on identifying the conserved  mass
 $M$ with the black hole {\it internal} energy  and the last two terms
with   the external work being  done {\it on} the black hole.

These identifications are standard lore, but suppose we now consider
 the
redefined mass
\be
{\tilde M}\; =\;M\;-\; \Phi Q \; - \; \Omega J \;,
\label{6}
\ee
then the  first law can just as easily be written in the equivalent
 form
\be
d{\tilde M}\;=\; TdS \;-\; Qd\Phi  \; - \; Jd\Omega  \;.
\label{7}
\ee
Note that the absolute value of the  last two terms now represents the
 work 
that is done {\it by} the black hole on its environment. 

\par
Although Eq.(\ref{5}) is the common-place form  of the first law 
for black holes,
one may ask if  it is any more (or less) legitimate than its
 alternative
representation in  Eq.(\ref{7}). Actually, the answer --- like
in most things scientific or otherwise ---  depends on the context.
For instance, suppose some adventurous scientist is monitoring
the energy fluctuations of a black hole under some
sort of (yet-to-be-elaborated-on) experimental procedure. If the
 experiment
was such that the charge and angular  momentum were to be held fixed,
 Eq.(\ref{5})
would be more appropriate for the  subsequent analysis. On the other
 hand, 
our hypothetical scientist may have
better control over the  electrostatic potential and rotational
 velocity
 (which, 
if anything,
 seems the more realistic scenario) and would, thereby, prefer to think in
 terms of
Eq.(\ref{7}). To put it another way, the physically motivated context
 should
 determine
 the
choice of extensive variables and, consequently, the most appropriate
 formulation
of the first law. 

\par
Continuing along these lines, let us take
notice of the relative signs of the work
terms in Eqs.(\ref{5}) and (\ref{7}). 
Comparing with text-book thermodynamic relations,
one can observe that the mass   $M$
is actually indicative of an ``enthalpic-type'' of potential, while
it is the quantity ${\tilde M}$
that is more suggestive
of an  internal energy. (Such an
observation, although in stronger terms, was recently made by
Shen {\it et al}
 \cite{cai}. 
Consult the cited paper
 for
further justification of this interpretation.)
To clarify  this last point, it is (in spite of many statements to
the contrary) not  {\it a priori} clear what
should be regarded as the definitive internal energy of a black hole.
That is to say, Hawking's calculation pinpoints the temperature,
which then vicariously fixes the  entropy given the functional
form that was  proposed by Bekenstein. After that, it is an open
 question
as to how the other terms  in the first law should be divided up
between work and internal energy. Really, only for the Schwarzschild
case ($Q=J=0$) can this division  be regarded as unambiguous. 

\par
What does all this mean? If {$\tilde M$} --- and not $M$ --- is
the
 true measure of  a black hole's internal energy,
it might be fair to
argue on behalf of Eq.(\ref{7}) as being the more
``fundamental'' of the two realizations. On the other hand, most
researchers would probably regard $Q$ and $J$ (rather than $\Phi$ and
$\Omega$) as being the ``natural'' choice  for the extensive work
 variables; 
thus implicating Eq.(\ref{5}) as the more fundamental statement. 
Either way,
the current author would rather allow the two
formulations to  have an equal status; with
 a particular
 choice being (as stressed above) predicated  on the situational 
or experimental context.~\footnote{As far as favoring ${\tilde M}$
over $M$ with
the assignment of
internal energy (as argued for in \cite{cai}), the current author
 chooses
 to remain agnostic. Hence, we will adopt the term ``alternative
energy" when  verbally referencing ${\tilde M}$.}

\par
Let us now slightly  alter  course and talk about  what is known as
Ruppeiner geometries  and, in particular, how these can relate
to black hole thermodynamics.  The story begins, some time ago, with 
Weinhold \cite{wein} proposing a  metrical structure and,
hence, a geometrical description for a given thermodynamic system.
 More specifically, the proposed
metric is based upon the  Hessian (or second-order partial derivatives)
 of the 
internal energy with respect to the entropy and any  other 
extensive variables of the system. 
A few years later, Ruppeiner \cite{rupe} made a similar proposal, but
 now
 with the 
metric
being defined in terms of the Hessian of the entropy.
To be more formal, let $U$ be the internal energy, $S$ be (as always)
 the 
entropy
and let  $X_i$ [$Y_i$] collectively denote {\it all} of the system's  
extensive variables 
{\it except} for $U$ [$S$]. Then
the Weinhold and Ruppeiner metrics can respectively be represented as
 follows:
\be
ds^2_W\;=\;+\partial_i\partial_jU\;dX_idX_j
\label{8}
\ee
and
\be
ds^2_R\;=\;-\partial_i\partial_jS\;dY_idY_j \;.
\label{9}
\ee

For future reference, as well  as a point of interest in its own right,
it can be shown that
\cite{conform-1,conform-2}
\be
ds^2_R\;=\; {1\over T} ds^2_W \;,
\label{10}
\ee
where $T=\partial_SM$ is (of course) the temperature.

\par
It is Ruppeiner's contention --- and convincingly supported
 \cite{rupe-2} 
--- that  Eq.(\ref{9}) 
describes a Riemannian
geometry which  provides a substantial amount of information
about the corresponding  thermodynamic system and
its statistical-mechanical counterpart.~\footnote{In this regard, the
 Weinhold
geometry  has some utility of its own, but it
will not be considered  here except as an
intermediary calculational tool.}
For instance, he has
demonstrated  that a flat-space metric indicates    
a non-interacting  statistical-mechanical system 
(with the converse having been validated as well). Moreover,  the resultant
scalar curvature has been shown to provide significant  information
about the system's  thermodynamic stability. In particular,
curvature 
singularities
are expected to be in one-to-one correspondence with critical or
 phase-transition
 points. Also, localized zeroes should indicate thermodynamically
special points where the interactions are ``turned off''.
\par
It is natural to apply   the Ruppeiner metrical formalism  
to black holes, given their current status as ``full-fledged''
 thermodynamic systems
(as discussed above~\footnote{In spite of our previous
discussion, such a claim is still somewhat open 
to debate \cite{helfer}.}). 
Indeed, much has been done in this field,
with Aman and collaborators at the forefront 
\cite{aman-1,aman-2,aman-3,aman-4,aman-5,aman-6}.
(For some other relevant work, see
 \cite{xxx-1,btz-rup,mirza,xxx-2}.)
However, even if black holes are thermodynamic entities in the truest
 sense,
there are both conceptual and technical  barriers that can impede
such an  application. Most notoriously, black hole thermodynamics
still lacks a {\it universally} acceptable statistical 
explanation.~\footnote{There are, however, model-specific explanations
 such as
(for instance) certain string-theoretic  extremal and near-extremal
 black holes 
\cite{peet}.}
 Furthermore, many black hole systems are dangerously unstable due
to negative heat capacities. And let us not overlook, even if
only recently having come to light \cite{cai}, 
what choice should be made for the internal energy?

\par
In view of such caveats, it should probably not be
 too 
surprising
that the Ruppeiner metric  can yield some
perplexing --- and perhaps even disturbing --- results in a black hole
context. 
Most infamously,
a Reissner--Nordstrom black hole ({\it i.e.,} a charged but
non-rotating black hole in an asymptotically flat spacetime)
produces a perfectly flat  Ruppeiner geometry 
\cite{aman-1}. (Importantly, this result persists for any 
applicable 
dimensionality \cite{aman-2} and for a family of closely related
 dimensionally
reduced models \cite{aman-4}.)
Aman {\it et al}  consistently argue 
that this  flatness should  not necessarily have gone unanticipated; 
rather,  they attribute the flat Ruppeiner metric to 
the scale invariance of the Einstein--Maxwell action \cite{aman-3}. 
The implication
then being that such scale invariance is (somehow) the black hole
 analogue
of a non-interacting statistical system. If
we put this rationalization aside, what is perhaps most discomforting
is that a flat-space metric  means (quite obviously) an
 everywhere-vanishing 
scalar curvature.
Hence, the curvature is rendered incapable of providing any information
about the interesting points of the Reissner--Nordstrom phase space;
namely,
the extremal point where the inner and outer horizon coalesce and 
the Davies critical point \cite{dav} where the heat capacity diverges.
(The status of the Davies point as a meaningful transition
point is highly controversial \cite{anti-dav}. 
Those who argue on behalf of the dissenting
side would most likely view the vanishing curvature as further support 
for their case. Still, it seems rather bothersome that a singularity
in the  formalism could be {\it completely} overlooked by
a function that aspires to provide {\it detailed}  information about
the system's stability.)

\par
Whether or not this  Ruppeiner flatness
(and any implications thereof)      
is a tenable state of affairs
would
be up to each individual reader to decide.
But, suffice it to say, there has been substantial backlash to this
 outcome in
the form of proposed ``resolutions''. That is, formal modifications
that are able to induce  
curved Ruppeiner metrics~\footnote{In the case of proposal {\it (iv)},
the metric is no longer Ruppeiner {\it per se}
but rather ``Ruppeiner-like''.}
for the Reissner--Nordstrom family after all. 
These approaches
 include the following: {\it (i)} adding
a physically motivated regulator 
(in the guise of a  spacetime curvature and/or 
an angular momentum)
that is  taken to zero only at the end of the calculation \cite{mirza},
{\it (ii)} taking quantum effects such as thermal fluctuations   
 into account  \cite{btz-rup}, {\it (iii)}
redefining the thermodynamic metric on the basis of invariance
under canonical transformations \cite{xxx-2} and, as already mentioned,
{\it (iv)}
using the extensive variables as prescribed by
the ``reformulated''  first law of Eq.(\ref{7}) \cite{cai}.

\par
Later on, we will comment briefly on the first three of
these proposals, 
but our focus will be  mainly on the fourth. In the next section, we
 will 
present two distinct calculations of the Ruppeiner metric and 
its associated scalar curvature. The distinction to be made is between
the two (presumably) valid choices for the
black hole internal energy; that is, $M$ and ${\tilde M}$.
This will allow  the reader a clear picture of how the
talked-about results actually
transpire. An interpretation and further discussion will then be
 provided in 
the final section.

\par
 For an analytic illustration that optimizes clarity  
(and so that no symbolic computing is required),
we will adopt the  rotating BTZ
 black hole \cite{btz-1,btz-2} as our system of 
choice.~\footnote{For an earlier study on the BTZ black hole
and thermodynamic geometries, see \cite{cai-2}.} 
In its most literal interpretation, the BTZ black hole  
is a ``toy model''
of a three-dimensional black hole in a (necessarily) negatively
curved spacetime. It mimics many of the features of higher-dimensional
(and, presumably, more realistic) black holes, while providing much
 more 
tractable calculations. But, in spite of this simplicity, the BTZ black
 hole
does play an important role in the near-horizon limit of many
 string-theoretic 
scenarios \cite{btz-string} and --- by virtue of the 
AdS--CFT correspondence \cite{ads-cft} ---
can be viewed as a holographic dual to an almost extremal 
Reissner--Nordstrom black hole \cite{btz-dual}. 
 
\section{Ruppeiner (BTZ)  Geometry}

\par
As a starting point for our analysis, it is useful to recall
the metric for a rotating BTZ (three-dimensional
anti-de Sitter) black hole \cite{btz-1}:
\be
ds^2_{BTZ}\;=\;-N^\perp dt^2 \;+\;{1\over N^\perp}dr^2\;
+\;r^2(N^{\phi}dt+d\phi)^2\;,
\label{11}
\ee
where
\be
N^{\perp} \;=\;{r^2\over l^2}\;-\;8GM\;+\;{J^2\over 4r^2}\;
\label{12}
\ee
and
\be
N^{\phi}\;=\;-{J\over 2r^2}\;.
\label{13}
\ee
Also, $G$ is the three-dimensional Newton's constant 
(necessarily being of dimension length),
$l$ is the radius of curvature (corresponding to a cosmological
 constant
of 
$-1/l^2$), $M$ is the conserved mass and $J$ is the conserved angular
momentum (which is physically equivalent to
a U(1) Abelian charge~\footnote{We will assume, without loss of 
generality, that $J$ is non-negative.}). 

Solving for the two zeros of $N^\perp$, 
which have significance
as an outer and inner horizon $r_{\pm}$ , one readily 
obtains~\footnote{If $J =  8GMl$, then the two horizons coincide
and we have the special case of an extremal black hole.
It should be kept in mind that, for this  case, the temperature is 
a vanishing quantity.
Further note that  $J > 8GMl $ would mean a naked singularity,
and so this parameter range
 is not part of the relevant (thermodynamic) phase space.}
\be
r^2_{\pm}\;=\;4GMl^2\left[\;1\;\pm\;\sqrt{1\;-\;\left({J\over
 8GMl}\right)^2}
\right]\; .
\label{14}
\ee
It is also useful to note that
\be
M \; =\; {r_{+}^2\;+\;r_{-}^2\over 8Gl^2} 
\;\;\; {\rm and}\;\;\; J \;=\;
2{r_{+}r_{-}\over l} \;. 
\label{15}
\ee

\par
Because this geometry is, locally, just  anti-de Sitter space,
there are many subtleties
to the BTZ framework \cite{btz-2}.
However, for our purposes, it is sufficient to remark
that the spacetime can --- with the appropriate identifications ---
 be interpreted as a black hole with an
outer horizon  having
the following thermodynamic  properties: an entropy
of $S=A/4G=\pi r_{+}/2G$, a  temperature of 
$T=\kappa/2\pi=\partial_rN^\perp(r=r_+)/4\pi$  and a rotational velocity of
$\Omega=r_{-}/r_{+}$.  
To avoid obscuring the calculations with needless clutter, we will 
subsequently set $8G=1$ and $l=1$.
Along the same lines,  Boltzmann's constant (previously 
set to unity) will now be calibrated to $k_B=1/4\pi$; a choice that
conveniently fixes 
$S=r_+$. 
  
\par
Our intent is to calculate the Ruppeiner scalar curvature
for the two previously discussed choices of  extensive variables. 
Firstly,  {\bf A.} the
``conventional''  set  $\lbrace$ $M$, $S$, $J$ $\rbrace$ which
 identifies
the conserved mass $M$ as the (black hole) internal energy
and then  {\bf B.} the ``unorthodox'' set   $\lbrace$ ${\tilde M}$, $S$,
 $\Omega$
$\rbrace$ for which 
\be
{\tilde M}\;=\;M\;-\;J\Omega
\label{16}
\ee
can also  be plausibly viewed, in accordance with our 
previous discussion, as the internal energy.

\par
It turns out that, rather than perform a direct calculation, it is 
much easier to 
compute the Weinhold metric
(\ref{8}) and then use the conformal transformation of Eq.(\ref{10})
to obtain the Ruppeiner metric (\ref{9}). We will proceed accordingly.

\subsection{$ \lbrace$ $M$, $S$, $J$ $\rbrace$ }

Here, the first step is to find the functional form $M=M(S,J)$.
Given  our choice of conventions (see above), it can readily be
 verified that
[{\it cf}, Eq.(\ref{15})]
\be
M\; =\; S^2 \;+\; {J^2\over 4S^2}\;.
\label{17}
\ee
By taking the second-order partial derivatives of $M$ with respect to
$S$ and $J$, one obtains the Hessian of the mass and
thus
\bea
ds_{W}^2 \;&=&\;\partial^2_SM\;dS^2\;+\;\left(\partial_J\partial_S
+\partial_S\partial_J\right)M\;dSdJ\;+\;\partial^2_JM\;dJ^2 
\nonumber \\
\;&=&\; \left(2+{3J^2\over 2S^4}\right)dS^2\;-\;2{J\over S^3}dSdJ
\;+\;{1\over 2S^2}dJ^2\;.
\label{18}
\eea

We can diagonalize the above metric by replacing $J$ with $x=J/S^2$.
This process yields
\be
ds_{W}^2\;=\; \left(2-{x^2\over 2}\right)dS^2\;+\;
{S^2\over 2}dx^2\;.
\label{19}
\ee

Next, let us consider the temperature in terms of $S$ and $x$:
\be
T\;=\;\partial_S M \;=\;2S\;-\;{J^2\over 2 S^3}\;=\;2S\;-\;{1\over
 2}Sx^2\;;
\label{20}
\ee
by which  Eq.(\ref{10}) prescribes the following for the Ruppeiner
 metric:
\be
ds_{R}^2\;=\; {1\over S}dS^2\;+\;
{S\over (4-x^2)}dx^2\;.
\label{21}
\ee

A further coordinate transformation to the variables  $y=2\sqrt{S}$
and $\omega$,  such that $x=2{\sin\omega}$, then  gives us
\be
ds_{R}^2\;=\; dy^2\;+\;
 y^2 d\omega^2\;.
\label{22}
\ee
This is easily recognizable as a flat Euclidean disc, meaning that
--- at least in this case ---
the Ruppeiner metric has an everywhere-vanishing 
curvature.~\footnote{This outcome is already known  \cite{aman-1}.
We have, however, documented the calculation for completeness
and illustrative purposes.}

\subsection{$ \lbrace$ ${\tilde M}$, $S$, $\Omega$ $\rbrace$ }

This time around,  it is the functional form of the alternative energy,
 ${\tilde M}={\tilde M}(S,\Omega)$, that is required.
By way of Eqs.(\ref{15}-\ref{16}) and $\Omega=r_{-}/r_{+}=J/2S^2$ 
(also keeping
our conventions in mind), it follows that
\be
{\tilde M}\;=\;S^2\left(1\;-\;\Omega^2\right)\;.
\label{23}
\ee
A calculation of the relevant Hessian ({\it i.e.}, ${\tilde M}$ varied
 by $S$
 and 
$\Omega$)  now leads to
\be
{\tilde ds}_{W}^2\;=\; 2\left(1-\Omega^2\right) dS^2\;-\;8S\Omega\; dSd\Omega
\;-\;2S^2d\Omega^2\;.
\label{24}
\ee

As before, we will transform coordinates so as to diagonalize the
metric. This can be achieved by eliminating $\Omega$ in favor of
 $X=\Omega S^2$,
and one then obtains
\be
{\tilde ds}_{W}^2\;=\; 2\left(3{X^2\over S^4}+1\right) dS^2
\;-\;{2\over S^2}dX^2\;.
\label{25}
\ee

In terms of $x$ and $S$, the temperature is expressible 
as~\footnote{As a consistency check, one can readily confirm the 
equivalency of our  various forms for  $T$; namely,
Eq.(\ref{20}), Eq.(\ref{26}) and [recalling that $k_B=1/4\pi$ ]
 $\partial_rN^\perp(r=r_+)$.} 
\be
T\;=\;\partial_S {\tilde M} \;=\;2S\left(1-\Omega^2\right)\;=
\;{2\over S^3}\left(S^4\;-\;X^2\right)\; ;
\label{26}
\ee
and so it follows that
\be
{\tilde ds}_{R}^2\;=\; {1\over S}{\left(3X^2+ S^4\right)\over \left(
S^4-X^2\right)}dS^2
\;-\;{S\over \left(S^4-X^2\right)}dX^2\;.
\label{27}
\ee

Given a two-dimensional diagonal metric of the generic
form $ds_g^2=-Ada^2+Bdb^2$ (with $A$ and $B$ both positive), 
the scalar curvature can be
computed by way of~\footnote{This equation can be obtained  from
the standard Riemannian-curvature formulae via the famed 
``brute-force'' method.}
\be
R_g={1\over 2A}\left[\partial^2_a \ln B + \left(\partial_a\ln
 B\right)^2
-\partial_a\ln A\partial_a\ln B\right]- 
{1\over 2B}\left[\partial^2_b \ln A + \left(\partial_b\ln A\right)^2
-\partial_b\ln A\partial_b\ln B\right]\;. 
\label{28}
\ee
Directly plugging in  the Ruppeiner  metric of Eq.(\ref{27}), one finds
(after some simplification) that
\be
{\tilde R}_{R}\;=\; 2{S^3\left(S^4-X^2\right)\over\left(S^4+3X^2\right)^2 }\;.
\label{29}
\ee
It is illustrative to re-express the above in terms of  more 
familiar parameters, such as the horizon radii,
\be
{\tilde R}_{R}\;=\;
 2{r_+\left(r_+^2-r_-^2\right)\over\left(r_+^2+3r_-^2\right)^2 }\;.
\label{30}
\ee

Unlike the prior (orthodox) case, we now have a decidedly
non-flat Ruppeiner geometry. Moreover, 
the curvature is of just such a form   that it provides us 
with a  clear signal for each one of
the anticipated pair  of thermodynamically ``special'' points.
Specifically, there is (first of all) the extremal point defined
by the coincidence of the horizons or $r_-=r_+\;$. At precisely
this point, we
observe a vanishing curvature; quite sensibly,
inasmuch as all of the degrees of
freedom should  be ``frozen'' at extremality [{\it cf}, Footnote 9].
  
Secondly, there happens to be  another type
of extremal point  --- or let us rather say "quasi-extremal"  point
---  associated with the BTZ black hole.  To elaborate, let us
begin by considering the near-horizon limit of a near-extremal 
Reissner--Nordstrom  black hole. After some degree of scrutiny, one finds
that the resultant geometry asymptotes to  that of a nearly massless
BTZ black hole \cite{btz-dual}.~\footnote{To be absolutely precise,
the Reissner--Nordstrom geometry asymptotes (in this limiting case)
to a dimensionally reduced version
 of the (near-massless) BTZ black hole \cite{btz-reduce,zzz}.}
Then, by virtue of this duality,
one could certainly  anticipate a  BTZ black hole  
to be  thermodynamically special at 
the point of vanishing mass.  
This expectation is, indeed,  realized
by the above  result; with  the Ruppeiner curvature ``blowing up'' 
as this limit of zero mass is approached 
(equivalently, ${\tilde R}_R\rightarrow \infty$
as $r_\pm\rightarrow 0$).~\footnote{It might be thought that
a divergence in this  limit is trivial for reasons of
dimensionality: $[R]\sim {\rm length}^{-2}\;$.
The reader should, however, keep in mind that
there is  another length scale in the problem; namely, the curvature
radius $l$.
Hence, it is a definitively non-trivial result that a massless
 black hole induces
a divergence in the curvature.} Interestingly, one could
argue that, from the perspective
of  the BTZ model, the Reissner--Nordstrom extremal point represents
some type of phase transition \cite{yyy}. 

\section{Interpretation}

To summarize the prior section, we have explicitly demonstrated (for
the rotating BTZ model) a remarkable result: Given a black hole
as a thermodynamic system, 
how one chooses to identify the extensive variables can have a very dramatic
 effect
on the induced Ruppeiner geometry. 
As anticipated, the ``conventional'' choice of
$ \lbrace$ $M$, $S$, $J$ $\rbrace$  results in a completely flat
metric with an everywhere-vanishing curvature. Conversely, the
``unorthodox'' set of variables $ \lbrace$ ${\tilde M}$, $S$, $\Omega$
 $\rbrace$
 induces a curved geometry that is undoubtedly reflective of
the ``special'' points of the thermodynamic system.

Before attempting an interpretation, it is useful to bring the
discussion back around  to the more-publicized Reissner--Nordstrom
 case.
Here, one similarly finds that \cite{aman-1}
\be
R_R=0 \;\;\; {\rm  everywhere}
\label{31}
\ee
 for the orthodox choice of 
$ \lbrace$ $M$, $S$, $Q$ $\rbrace$ .  Meanwhile,
a set of variables based on the  ``alternative" (internal) energy 
 $ \lbrace$ ${\tilde M}$, $S$, $\Phi$ $\rbrace$ 
will yield the following \cite{cai} (up to  an irrelevant
numerical factor):
\be
{\tilde R}_{R}\;\sim\; {r_+-r_- \over r_+\left(r_+-3r_-\right)^2 }\;.
\label{32}
\ee
It should be noted that there is a  zero in Eq.(\ref{32})
at the extremal point $r_-=r_+$ and a singularity at the
so-called Davies point \cite{dav} $r_+=3r_-$. The former is 
thermodynamically special
because of a vanishing temperature, while the latter corresponds
to a divergent heat capacity (and, at least naively, 
a transition in phase). Let us again remind the reader that
the status of the Davies point  as a ``legitimate'' critical
 point
is notably controversial \cite{anti-dav}; this,  in spite of the 
singular  specific heat.

So what exactly  do we have here? The former (conventional) choice
of extensive variables
tells us that the underlying system is trivially simple: whatever, the
 black
hole analogue of ``non-interacting'' means. (According to Aman {\it et
 al},
this means the scale invariance of the Einstein--Maxwell action
 \cite{aman-3}.)
On the other hand, the latter (unorthodox) choice induces
a curved geometry that is indicative of a highly interacting
system.  What is more, the pole in the  scalar curvature legitimizes 
the status of the Davies point. [It is interesting to note that, when
 taken
together, Eqs.(\ref{31}) and (\ref{32}) serve to both encapsulate and
perpetuate the very controversy this formalism should be resolving.]
At the very least, we appear to have a disturbing contradiction.

But let us  not be so quick to rush to
a despairing judgment. The prominent mantra of 
the introductory section is that the correct choice of energy
($M$ or ${\tilde M}$) should be dependent on the particular situation
 at
 hand; 
that is,
the choice should really depend on what type of physical question 
is being posed.
If this is a valid assessment, then it would be fair to say that
each of the above results could have validity within a  certain
 context. On
this basis, let us conjecture as follows: If we want to
know about the underlying statistical system, then  
$M$ should be selected as the  energy, which leads us to the
 flat-curvature
result of Eq.(\ref{31}). However, if we  want to know about the
 thermodynamic
phase space, then ${\tilde M}$ is the best choice for the energy,
thus arriving at the non-trivial geometric picture of Eq.(\ref{32}).

To elaborate, it is difficult to say anything definitive about the
 statistical
framework underlying the thermodynamics of a Reissner--Nordstrom black
hole. Unlike asymptotically anti-de Sitter spacetimes, there is
no convenient duality ({\it vis-a-vis}, the AdS--CFT correspondence 
\cite{ads-cft})
to ``hang one's hat on''. Given that a pure Reissner--Nordstrom black
 hole
is an idealized fiction (the uncertainty principle
forbids  an exactly  vanishing angular momentum and, depending
on the exact origin  of the dark energy,  possibly
a vanishing cosmological constant as well),  it is quite feasible that
  the statistical origin of its thermodynamics is either trivial
or even non-existent. Perhaps, this is precisely what Eq.(\ref{31})
is trying to tell us!
 
But, irrespective of the statistical mechanics, it is clear that
eventful things take place in the thermodynamic phase space
of a Reissner--Nordstrom black hole. The temperature can certainly
vanish (at the extremal point), the heat capacity can diverge
(at the Davies point) and both of these quantities (as
well as many others) can vary greatly as the extensive parameters
are being tuned. We would suggest these to be the very type
of features that Eq.(\ref{32})
is  telling us about! 

The inquiring reader might now ask: does this interpretation mean the
Davies point is a legitimate phase transition point after all?
In our opinion, the answer is a qualified no. We reiterate that,
{\it standing on its own}, the Reissner-Nordstrom black hole
is a toy model with no real validity as a thermodynamic system.
After all, the temperature is calculated on the basis of
a quantum process ({\it i.e.}, the  Hawking Radiation effect
 \cite{hawk}),
and, once  quantum mechanics enters into the discussion, then we must
 allow
for fluctuations in all other parameters --- including the angular
momentum and anything else of pertinence (such
as any relevant `` quantum hair''). 
Meaning, in a truly realistic
system, it is quite feasible  that the Davies point is elevated to
a phase-transition point;  assuming a point 
of divergent heat capacity survives  at all  --- this is certainly
 not guaranteed.  

Let us now  briefly comment upon some other proposals for
``resolving'' the flat  Ruppeiner geometry for the Reissner--Nordstrom
class of black holes. Let us first  consider the work of Mirza and
Zamani-Nasab \cite{mirza}. These authors regulate the
 Reissner--Nordstrom 
calculation by 
starting with a ``fully loaded'' Kerr--Newman--AdS black 
hole~\footnote{That is, a black hole inside of
an asymptotically (negatively) curved spacetime and having, in general,
a non-vanishing charge and momentum.
The Reissner--Nordstrom black hole can, in some sense,
  be regarded as a special case of
 this more general model.} and then return
to the Reissner--Nordstrom case only at the very end of the
 calculation.
In this way, they manage to obtain a curved Ruppeiner metric in
what can be argued as being a more physically motivated calculation.  
Also with the intent of steering in the direction of  realism, 
Sarkar {\it et al} \cite{btz-rup} consider some quantum-inspired
modifications to the BTZ model.
For instance, these authors find that the inclusion of thermal
fluctuations  does indeed induce a curved Ruppeiner metric.
Both of these 
findings are certainly {\it copacetic} with some of the comments made
 above. 

Very recently, Alvarez {\it et al} 
\cite{xxx-2}~\footnote{So recent, in fact, that this paper was
 released on
to the 
archives while the current
paper was   in a preparatory phase.}
 more directly address the $M$ versus
${\tilde M}$ debate. They derive a Ruppeiner-like  metric that is based
 on 
the principle
of invariance under canonical transformations. (Meaning, that it is
 independent
of the choice made for the extensive variables.) Their final outcome
for the Reissner--Nordstrom black hole is qualitatively very similar
 to that of Eq.(\ref{32}) and certainly nothing like that of
 Eq.(\ref{31}).
We would, however, question whether their formalism can be viewed as a 
replacement --- as they strongly imply~\footnote{To quote the authors
on their own findings: ``This result finishes the controversy regarding
the application of geometric structures in black hole thermodynamics'' 
\cite{xxx-2}.} --- as opposed to an alternative to
the Ruppeiner metric. Moreover, their result --- if taken in isolation
 ---
would miss out completely on  whatever lesson is to be learned from
the flat-space metric of Eq.(\ref{31}). Like detectives assembling
information during a crime investigation, we suggest that
any pertinent resource should be utilized before reaching a conclusion.

Assuming there is some degree of  validity to our interpretations and
 overview, 
the next logical step
would be to understand this critical distinction that arises
between the two choices of extensive variables.
 That is, why does the conventional choice and the unorthodox choice
 lead,
respectively, to
Eq.(\ref{31}) and Eq.(\ref{32}),  and not {\it vice versa}?
Or to put it yet another way, why is the conserved mass $M$ more
 sensitive
to the statistical mechanics   and the alternative energy 
${\tilde M}$, to the thermodynamics of the black hole system?
We hope to address this issue in due course.

\newpage
\section*{Acknowledgments}
Research is financially supported by the University of Seoul.
The author thanks Yun Soo Myung for
his inputs and valued discussions, and CQUeST  at Sogang University
for their hospitality.


\end{document}